\documentclass[hyper]{JHEP3} 

\usepackage{epsfig,multicol}

\newcommand\fverb{\setbox\pippobox=\hbox\bgroup\verb}
\newcommand\fverbdo{\egroup\medskip\noindent%
                        \fbox{\unhbox\pippobox}\ }
\newcommand\fverbit{\egroup\item[\fbox{\unhbox\pippobox}]}
\newbox\pippobox

\def\bea{\begin{eqnarray}}
\def\eea{\end{eqnarray}}
\def\bec{\begin{center}}
\def\ec{\end{center}}

\def\beq{\begin{equation}}
\def\eeq{\end{equation}}

\title{String theoretic QCD axion with stabilized saxion
                  and the pattern of supersymmetry breaking}

\author{ Kiwoon Choi and Kwang Sik Jeong \\
         Department of Physics, Korea Advanced Institute of Science
         and Technology\\
         Daejeon 305-701, Korea\\
         E-mail: \email{kchoi@hep.kaist.ac.kr},
                 \email{ksjeong@hep.kaist.ac.kr} }

\preprint{KAIST-TH 2006/11}

\abstract{
String theoretic axion is a prime candidate for the QCD axion solving the
strong CP problem.
For a successful realization of the QCD axion in string theory, one needs
to stabilize moduli including the scalar partner (saxion) of the QCD axion,
while keeping the QCD axion unfixed until the low energy QCD instanton
effects are turned on.  We note that a simple generalization of KKLT
moduli stabilization provides such set-up realizing the axion solution to
the strong CP problem.
Although some details of moduli stabilization are different from the
original KKLT scenario, this set-up leads to the mirage mediation pattern
of soft SUSY breaking terms as in the KKLT case, preserving flavor and CP
as a consequence of approximate scaling and axionic shift symmetries.
The set-up also gives an interesting pattern of moduli masses which might
avoid the cosmological moduli, gravitino and axion problems.
}


\begin{document}

\section{Introduction}

The strong CP problem \cite{strongcp} is a naturalness problem arising from
that CP is conserved by the strong interactions but not by the weak
interactions.
The low energy QCD lagrangian contains a CP violating angle
$\bar\theta=\theta_{QCD}+{\rm arg}{\rm Det}(\lambda_u\lambda_d)+...$, where
$\theta_{QCD}$ is the bare QCD vacuum angle,
$\lambda_{u,d}$ are the Yukawa couplings of the up and down-type quarks,
and the ellipses stands for the contribution from other high energy
parameters, e.g. the gluino mass and $B$-parameter for the case of
supersymmetric models.
The observed CP violations in $K$ and $B$ meson system suggest that
$\lambda_{u,d}$ are complex parameters with phases of order unity.
On the other hand, the non-observation of the neutron electric dipole moment
implies that $|\bar{\theta}|\lesssim 10^{-9}$.
This raises the question why the phase combination $\bar\theta$ is so small.

There are presently three known solutions to the strong CP problem.
One simple possibility is that the up quark is massless,
rendering the CP violations from $\bar\theta$ vanish.
A massless up quark might not be in conflict with the known low energy
properties of QCD since an effective up quark mass can be mimicked
by instanton effects \cite{mup}.
A second solution is that CP is an exact symmetry of the underlying
fundamental theory \cite{gaugecp}, but is broken spontaneously in a specific
manner to give $\bar{\theta}$ small enough \cite{nelsonbarr}.
The third solution is to have a non-linearly realized global $U(1)_{PQ}$
symmetry which is explicitly broken by the QCD anomaly \cite{pq}.
This solution predicts a light pseudo Goldston boson, the axion \cite{axion},
which might have interesting cosmological and/or astrophysical implications
\cite{strongcp}.

Compactified string theory contains numerous axions which originate from
higher-dimensional antisymmetric tensor fields \cite{green}, thus is perhaps
the most plausible framework to give the QCD axion solving the strong
CP problem \cite{choikim,dine,witten1,conlon}.
If we assume supersymmetric compactification, axions are accompanied by
their scalar partners.
In this case, much of the physical properties of the QCD axion depends on the
mechanism stabilizing its scalar partner ``saxion".
For instance, the axion decay constant and the strength of unwanted
(non-perturbative) $U(1)_{PQ}$ breaking other than the QCD anomaly can be
determined only after the saxion vacuum value is fixed.
Axion cosmology is another subject depending severely on the saxion
stabilization mechanism.
Thus an explicit realization of saxion stabilization is mandatory in order to
see if a specific string compactification can successfully realize the axion
solution to the strong CP problem.

A key requirement for the saxion stabilization is that it should keep the
QCD axion as a flat direction until the low energy QCD instanton effects are
taken into account.
Unless the QCD axion mass from saxion stabilization  is extremely suppressed
as Eq. (\ref{bound}), the dynamical relaxation of $\bar\theta$ can not be
accomplished.
In light of the recent progress in moduli stabilization
\cite{fluxcompactification}, an immediate step toward string theoretic QCD
axion would be $U(1)_{PQ}$-invariant generalization of KKLT moduli
stabilization which starts with supersymmetric AdS solution lifted later to
dS (or Minkowski) vacuum by an uplifting potential \cite{kklt}.
In regard to this possibility, it has been noticed  recently that
supersymmetric solution of any $U(1)_{PQ}$-invariant effective SUGRA gives
a tachyonic saxion mass \cite{conlon}.
This might be considered as an indication that QCD axion favors
non-supersymmetric moduli stabilization such as the perturbative
stabilization discussed in \cite{hebecker} or the large volume stabilization
advocated in \cite{quevedo}.

In this paper, we point out that an uplifting potential induced by SUSY
breaking brane stabilized at the end of warped throat \cite{kklt,choi1,russel},
which is in fact the most plausible form of uplifting potential in KKLT-type
compactification,
automatically solves the tachyonic saxion problem for $U(1)_{PQ}$-invariant
generalization of KKLT moduli stabilization with the number of K\"ahler
moduli $h_{1,1}>1$.
We also examine the pattern of moduli masses and soft SUSY breaking terms of
visible fields in this set-up giving the QCD axion.

Quite interestingly, although some details of  moduli stabilization are
different from the original KKLT scenario, the resulting soft SUSY breaking
terms still receive comparable contributions from moduli mediation
(including the saxion mediation) \cite{modulimediation} and anomaly mediation
\cite{anomalymediation}, thereby take the mirage mediation pattern
\cite{choi1,mirage,mirage1,mirage2,mirage3,pierce} as in the KKLT
case\footnote{
We note that our saxion stabilization scheme does
not give the pattern of soft terms proposed  in \cite{conlon1}.
}.
Furthermore, the soft terms naturally preserve flavor and CP as a consequence
of approximate scaling and axionic-shift symmetries of the underlying string
compactification, independently of the detailed forms of moduli K\"ahler
potential and matter K\"ahler metric.
As for the flavor conservation, the universality of moduli (including the
saxion) $F$-components which is another interesting feature of our set-up
plays an important role.

Our moduli stabilization set-up gives also an interesting pattern of moduli
masses.
Independently of the detailed form of the moduli K\"ahler potential, saxion
has a mass $m_s\simeq \sqrt{2}m_{3/2}$, while the other K\"ahler moduli
(except for the QCD axion) have a mass of the order of
$m_{3/2}\ln(M_{Pl}/m_{3/2})$, and the visible sector superparticles have
soft masses of the order of $\frac{m_{3/2}}{\ln(M_{Pl}/m_{3/2})}$.
If the visible sector superparticles are assumed to have the weak scale
masses, the saxion has a right mass to decay right before the big-bang
nucleosynthesis (BBN), while the other moduli  are heavy enough to decay well
before the BBN. This feature leads to moduli cosmology different from the
original KKLT set-up \cite{kkltcosmology}, and might allow to avoid the
cosmological gravitino, moduli and axion problems \cite{choi4}.
In particular, it might allow the QCD axion to be a good dark matter
candidate under a mild assumption on the initial axion misalignment although
the axion decay constant is near the GUT scale.

\section{Saxion  stabilization}

Let $T$ denote the modulus superfield whose pseudoscalar component
${\rm Im}(T)$ corresponds to the QCD axion solving the strong CP problem.
For the dynamical relaxation of $\bar\theta$, ${\rm Im}(T)$ is required to
couple to the QCD anomaly $F\tilde{F}$, i.e. the holomorphic gauge kinetic
function of QCD should depend on $T$ as
\bea
\label{gaugekinetic}
f_a = c_TT+\Delta f_a(\Phi^i),
\eea
where $c_T$ is a real nonzero constant, and $\Phi^i$ are generic moduli
other than $T$.
To avoid saxion-mediated macroscopic force, the saxion
$s=\sqrt{2}{\rm Re}(T)$ should be stabilized with $m_s\gtrsim 10^{-3}$ eV.
In fact, cosmological consideration typically requires  much heavier
saxion mass, e.g. $m_s\gtrsim 40$ TeV for the saxion decay before the BBN
in case that saxion couplings are Planck-scale suppressed \cite{casas}.
On the other hand, in order to keep the dynamical relaxation of $\bar\theta$
available, the axion  $a=\sqrt{2}{\rm Im}(T)$ should remain to be unfixed
until the low energy QCD instanton effects are taken into account.
More explicitly, saxion stabilization mechanism should preserve the
non-linear PQ symmetry
\bea
\label{u1pq}
U(1)_{PQ}: \,\,T\rightarrow T+i\beta \quad
(\beta=\mbox{real constant})
\eea
to the accuracy that axion mass induced by the saxion stabilization is small
as
\bea
\label{bound}
\delta m_a\,\lesssim\, \frac{\sqrt{10^{-9}m_\pi^2
f^2_\pi}}{v_{\rm PQ}}\,\sim\, 10^{-6}\left(\frac{10^9\,{\rm
GeV}}{v_{\rm PQ}}\right) \,{\rm eV},
\eea
where $v_{\rm PQ}$ is the axion decay constant which is constrained to be
bigger than $10^9$ GeV \cite{strongcp}.

The above condition strongly suggests that saxion should be stabilized within
the framework of $U(1)_{PQ}$-invariant effective SUGRA in which the K\"ahler
potential and superpotential take the form
\bea
\label{invariantsugra}
K = K(T+T^*,\Phi^i,\Phi^{i*}),
\quad W=W(\Phi^i).
\eea
Note that a $T$-dependent superpotential generically breaks
$U(1)_{PQ}$\footnote{
A superpotential of the form
$W=e^{-bT}\Omega(\Phi^i)$ ($b=\mbox{real constant}$) preserves $U(1)_{PQ}$.
However such form of superpotential typically leads to a runaway of saxion
unless an uncontrollably large quantum correction to K\"ahler potential is
assumed \cite{banks}.
Here we are interested in the possibility to stabilize the saxion in a region
of moduli space where the leading order K\"ahler potential is reliable.
},
giving an axion mass comparable to the saxion mass. A simple example of
$U(1)_{PQ}$-invariant saxion stabilization has been discussed before
\cite{hebecker} within the framework of flux compactification of type IIB
string theory. In \cite{hebecker}, it was assumed that the IIB dilaton and
complex structure moduli are stabilized by 3-form fluxes, leaving the single
K\"ahler modulus $T$ unfixed.
If the visible gauge fields live on $D7$ branes wrapping the 4-cycle whose
volume corresponds to ${\rm Re}(T)$, the QCD gauge kinetic function is given
by $f_a=T$, thereby ${\rm Im}(T)$ can be a candidate for the QCD axion.
At leading order, the K\"ahler potential of $T$ takes the no-scale form.
However at higher order in $\alpha^\prime$ and string loop expansion,
it receives $U(1)_{PQ}$-invariant corrections as
\bea
\label{kahler1}
K = -3\ln(T+T^*)+\frac{\xi_1}{(T+T^*)^{3/2}}-\frac{\xi_2}{(T+T^*)^2},
\eea
where $\xi_1$ is the coefficient of higher order $\alpha^\prime$ correction
which is positive for a positive Euler number, and $\xi_2$ is the coefficient
of string loop correction.
Assuming a flux-induced constant superpotential $W=w_0$, the resulting scalar
potential stabilizes ${\rm Re}(T)$ while keeping ${\rm Im}(T)$ unfixed if
$\xi_1>0$ and $\xi_2>0$.
In this scenario, saxion is stabilized by the competition between two
controllably small perturbative corrections, which is possible because the
potential is flat in the limit $\xi_1=\xi_2=0$.

Although attractive, the above saxion stabilization can be applied only for
$\xi_1>0$ which requires the Euler number $\chi=2(h_{1,1}-h_{2,1})>0$.
On the other hand, most of interesting Calabi-Yau (CY) compactifications
have nonzero $h_{2,1}$.
In particular, if one wishes to get a landscape of flux vacua which might
contain a state with nearly vanishing cosmological constant \cite{weinberg},
one typically needs  large number of 3-cycles, e.g.
$h_{2,1}={\cal O}(100)$, to accommodate 3-form fluxes
\cite{fluxcompactification}.
In such case, there remain (many) K\"ahler moduli not stabilized by the
above purely perturbative mechanism.

A simple way out of this difficulty would be to stabilize all K\"ahler
moduli other than saxion by non-perturbative superpotential a la KKLT,
while keeping the saxion stabilized by $U(1)_{PQ}$-invariant K\"ahler
potential.
In this case, the saxion potential resulting from the leading order K\"ahler
potential is {\it not} flat anymore,
therefore saxion can not be stabilized by controllably small perturbative
corrections to the K\"ahler potential.
Still this kind of generalized KKLT set-up might allow a supersymmetric
solution of
\bea
D_IW = \partial_IW+W\partial_IK=0
\eea
in a region of moduli space where the leading order K\"ahler potential is
reliable.
To see that this is a rather plausible possibility, let us consider a simple
example with
\bea
\label{toy}
K &=& -2\ln\left[
(T_1+T_1^*)^{3/2}-(T_2+T_2^*)^{3/2}-(T_3+T_3^*)^{3/2}\right],
\nonumber \\
W &=& w_0+A_1e^{-b_1T_1}+A_2e^{-b_2(T_2+T_3)},
\eea
where $w_0\sim m_{3/2}$ and $A_{1,2}\sim 1$ in the unit with $M_{Pl}=1$.
(Unless specified, we use the unit with $M_{Pl}=1$ throughout this paper.)
Here $K=-2\ln(V_{CY})$ corresponds to the leading order K\"ahler potential
of the K\"ahler moduli $T_i$ for a CY volume given by
\bea
V_{CY} = \int J\wedge J\wedge J=t_1^3-t_2^3-t_3^3,
\eea
where
$J$ is the K\"ahler two form and
$3(T_i+T_i^*)={\partial V_{CY}}/{\partial t_i}$.
For the above non-perturbative superpotential, it is convenient to define
new chiral superfields as
\bea
\Phi_1 = T_1, \quad \Phi_2=T_2+T_3, \quad T=T_2-T_3,
\eea
where $T$ corresponds to the invariant direction of $W$.
It is then straightforward to see that the model allows a SUSY solution:
\bea
T_1 &\simeq& \frac{1}{b_1}\ln(M_{Pl}/m_{3/2}),
\nonumber \\
T_2 &=& T_3\,\simeq \, \frac{1}{2b_2}\ln (M_{Pl}/m_{3/2}),
\eea
for which the leading order K\"ahler potential is a good approximation
if $m_{3/2}$ is hierarchically lighter than $M_{Pl}$.

Since it is always a stationary point of the scalar potential
\bea
V_{F} =
e^K\left(K^{I\bar{J}}D_IW(D_JW)^*-3|W|^2\right),
\eea
supersymmetric moduli configuration is a good starting point for moduli
stabilization.
On the other hand, it has been noticed that supersymmetric moduli
stabilization in $U(1)_{PQ}$-invariant effective SUGRA
(\ref{invariantsugra}) always gives a tachyonic saxion \cite{conlon}.
For $D_IW=0$  with $W\neq 0$, one easily finds
\bea
\label{modulipotential}
\left(\frac{\partial^2 V_{\rm F}}{\partial s^2}\right)_{D_IW=0}
&=& -2|m_{3/2}|^2\partial_T\partial_{\bar{T}}K < 0,
\eea
where $s=\sqrt{2}{\rm Re}(T)$ is the saxion and $m_{3/2}=e^{K/2}W$ is the
gravitino mass, thus there is a tachyonic direction which has a nonzero
mixing with the saxion field.
However supersymmetric moduli configuration generically gives an AdS vacuum,
thus requires an uplifting potential in order to be a phenomenologically
viable dS (or Minkowski) vacuum.
A simple way to get uplifting is to introduce SUSY breaking brane carrying
a positive tension as in KKLT \cite{kklt}.
If the underlying geometry has a warped throat \cite{rs,gkp}, the SUSY
breaking brane is stabilized at the end of throat \cite{kklt} independently
of the details of SUSY breaking dynamics.
In the following, we show that the uplifting potential induced by SUSY
breaking brane stabilized at the end of warped throat, which is the most
natural form of uplifting potential in KKLT-type moduli stabilization,
automatically solves the tachyonic saxion problem.

To this end, let us consider the KKLT-type compactification of Type IIB
string theory on CY orientifold with the number of K\"ahler moduli
$h_{1,1}>1$ and the visible gauge fields living on $D7$
branes\footnote{
If the visible gauge fields live on $D3$, the corresponding gauge kinetic
function $f_a=S$ can not give the QCD axion since the IIB dilaton $S$ is
fixed by the flux-induced superpotential
$W_{\rm flux}=\int \Omega\wedge (F_3-iSH_3)$.
}.
As usual, we assume that the string dilaton and all complex structure moduli
are fixed by 3-form fluxes with  masses hierarchically heavier than the
K\"ahler moduli and gravitino masses, and consider the effective SUGRA of
K\"ahler moduli
\bea
\Phi^I \,=\, (T,\Phi^i)=\frac{1}{\sqrt{2}}(s+ia,\phi^i+ia^i),
\eea
where ${\rm Re}(\Phi^I)$ and ${\rm Im}(\Phi^I)$ correspond to appropriate
linear combinations of the 4-cycle volumes and the RR 4-form axions,
respectively.
If ${\rm Re}(T)$ and $\Phi^i$ can be stabilized while preserving the
non-linear $U(1)_{PQ}$ symmetry (\ref{u1pq}), ${\rm Im}(T)$ can play the
role of QCD axion.
Following KKLT, we also assume that $\Phi^i$ are stabilized by
non-perturbative effects such as $D3$-instanton or a gaugino condensation on
the hidden $D7$-branes warpping certain 4-cycles in CY.
Finally, we introduce a SUSY breaking brane without specifying the SUSY
breaking dynamics, which will be stabilized at the end of (maximally) warped
throat independently of the detailed SUSY breaking dynamics.

After integrating out the heavy dilaton and complex structure moduli as well
as the SUSY-breaking fields on the uplifting brane, the effective action of
K\"ahler moduli can be written as \cite{choi1}
\bea
\label{Action}
&& \int d^4\theta CC^*\Big[\,-3e^{-K/3}
-CC^*\Lambda^2\bar{\Lambda}^2e^{4A}{\cal P}\,\Big]
\nonumber \\
&& \quad\qquad +\,\left[\int d^2\theta C^3\Big(w_0+\sum_i A_i
e^{-b_i\Phi^i}\Big)+{\rm h.c.}\right],
\eea
where $K$ is the effective K\"ahler potential of K\"ahler moduli
$\Phi^I=(T,\Phi^i)$, $C$ is the chiral compensator superfield of 4D $N=1$
SUGRA, $e^{4A}$ is the red-shift factor \cite{rs} of the effective
Volkov-Akulov action of the Goldstino superfield,
\bea
\Lambda^\alpha
= \theta^\alpha+\mbox{Goldstino-dependent terms},
\eea
which is localized on SUSY-breaking brane. This Volkov-Akulov action provides
a low energy description of generic SUSY-breaking brane stabilized at the end
of warped throat, e.g. anti-$D3$ brane or any brane which carries 4D dynamics
breaking $N=1$ SUSY spontaneously\footnote{
For more detailed discussion of this point, see
\cite{nilles,pierce,choi5,hebecker1,kobayashi}.
}.
Here the compensator dependence of the Volkov-Akulov term is determined by
that it corresponds to an uplifting potential with mass-dimension four in the
unitary gauge $\Lambda^\alpha=\theta^\alpha$, and the warp factor
$e^{4A}\ll 1$ is determined by the complex structure modulus of the
collapsing 3-cycle \cite{gkp}.
The axionic shift symmetries of the RR 4-form axions ${\rm Im}(\Phi^i)$,
\bea
U(1)_i:\, \Phi^i\rightarrow \Phi^i+i\beta_i \quad
(\beta_i=\mbox{real constants}),
\eea
are assumed to be broken by $D3$ instantons and/or hidden $D7$ gaugino
condensations generating the non-perturbative superpotential
$\sum_i A_ie^{-b_i\Phi^i}$,  while the axionic shift symmetry (\ref{u1pq})
for the QCD axion ${\rm Im}(T)$ is preserved at this stage.
In this scheme, possible non-perturbative $U(1)_i$-breaking in K\"ahler
potential can be ignored, and then $K$ takes the form:
\bea
K = K(T+T^*,\Phi^i+\Phi^{i*}).
\eea
Note that the chiral superfields $\Phi^i$ are defined through the exponents
of non-perturbative superpotential.
The constants $w_0$ and $A_i$ in the superpotential can be always made to be
real by appropriate $U(1)_R$ and $U(1)_i$ transformations, which will be
crucial for the soft SUSY breaking terms to preserve CP \cite{choi2}.

\begin{figure}[t]
\begin{center}
\begin{minipage}{15cm}
\centerline{
{
\hspace*{0.5cm}
\epsfig{figure=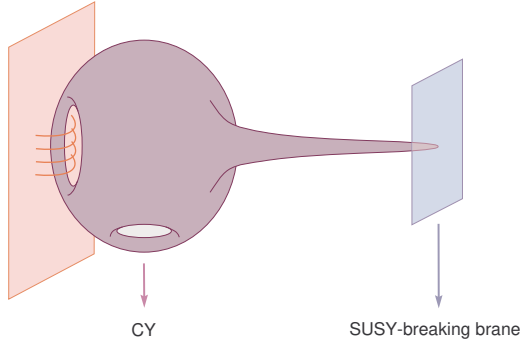,angle=-90,width=7cm}} }
\caption{ CY space and SUSY-breaking brane sequestered from each
other by warped throat. Here the visible sector $D7$ branes are
assumed to be wrapping a 4-cycle of CY. \label{fig:CY} }
\end{minipage}
\end{center}
\end{figure}

An important feature of KKLT-type compactification is that in the limit
$e^{4A}\ll 1/V_{CY}^{2/3}$, the uplifting brane is separated from CY by a
strongly warped throat: CY corresponds to the UV end of warped throat,
while SUSY-breaking brane is located at the IR end (See Fig. 1.).
In this case, K\"ahler moduli on CY and Goldstino superfield on SUSY breaking
brane are sequestered from each other \cite{choi1,sequestering,hebecker2},
i.e. ${\cal P}$ is independent of $\Phi^I$.
Indeed, a full 10-dimensional analysis gives \cite{giddings}
\bea
{\cal P} \,=\,
{\cal P}_0\Big[\,1+ {\cal O}(e^{4A}V_{CY}^{2/3})\,\Big]
\eea
for $e^{4A}\ll 1/V_{CY}^{2/3}$, where ${\cal P}_0$ is a $\Phi^I$-independent
constant\footnote{
Schematically, ${\cal P}$ depends on the K\"ahler moduli as
${\cal P}\propto 1/(1+e^{4A}V_{CY}^{2/3})$ \cite{giddings}, which gives
${\cal P}={\cal P}_0V_{CY}^{-2/3}[\,1+{\cal O}(1/e^{4A}V_{CY}^{2/3})\,]$
in the opposite limit of large volume and weakly warped throat with
$1/V_{CY}^{2/3}\ll e^{4A}$.
}.
Note that the visible sector $D7$ branes wrap a 4-cycle in CY at the UV end
of throat, thus we can realize the conventional high scale gauge coupling
unification even in the presence of highly warped throat.

From the above discussion, one finds that generic SUSY-breaking brane
stabilized at the IR end of warped throat provides a {\it sequestered} form
of Volkov-Akulov operator in $N=1$ superspace:
\bea
C^2C^{*2}\Lambda^2\bar{\Lambda}^2e^{4A}{\cal P}_0 =
C^2C^{*2}\theta^2\bar{\theta}^2e^{4A}{\cal P}_0
+ \mbox{Goldstino dependent terms},
\eea
where $e^{4A}{\cal P}_0$ is a constant.
In the Einstein frame with $C=e^{K/6}$, this Volkov-Akulov operator adds an
uplifting potential $V_{\rm lift}=e^{4A}{\cal P}_0e^{2K/3}$ to the
conventional $F$-term potential $V_F$, thereby the total moduli potential is
given by
\bea
V_{\rm TOT}=V_F+V_{\rm lift}=
e^K\Big(K^{I\bar{J}}D_IW(D_JW)^*-3|W|^2\Big)+e^{4A}{\cal P}_0e^{2K/3}.
\eea
On the other hand, the Volkov-Akulov operator does not affect the on-shell
expression of the moduli $F$-components, thus $F^I$ in the Einstein frame
takes the conventional form:
\bea
F^I=-e^{K/2}K^{I\bar{J}}(D_JW)^*.
\eea
Obviously, the uplifting potential $V_{\rm lift}=e^{4A}{\cal P}_0e^{2K/3}$
gives a positive saxion mass-square for supersymmetric moduli configuration
\bea
\left( \frac{\partial^2V_{\rm lift}}{\partial s^2}\right)_{D_IW=0}=\,
\frac{4}{3}V_{\rm lift}\partial_T\partial_{\bar{T}}K
\,\simeq\,
4|m_{3/2}|^2\partial_T\partial_{\bar{T}}K,
\eea
where we have used the SUSY condition $\partial_TK=0$ and also the condition
of vanishing cosmological constant:
\bea
\Big(V_{\rm lift}\Big)_{D_IW=0}\simeq
-\Big(V_F\Big)_{D_IW=0} \,\simeq\, 3|m_{3/2}|^2.
\eea
This positive saxion mass-square from $V_{\rm lift}$ always dominates over
the tachyonic saxion mass-square from $V_F$, thereby stabilizing the saxion
as
\bea
\left(\frac{\partial^2 V_{\rm TOT}}{\partial s^2}\right)_{D_IW=0}
\simeq
2|m_{3/2}|^2\partial_T\partial_{\bar{T}}K.
\eea

Schematically, $\Phi^i$ are stabilized by the KKLT-type of superpotential,
while the saxion is stabilized by $V_{\rm lift}$.
In this procedure, it is essential to have non-zero K\"ahler mixing between
$\Phi^i$ and $T$, i.e. $\partial_i\partial_{\bar{T}}K\neq 0$, as it allows
$\partial_TK=0$ in a region where the leading order K\"ahler potential is
reliable.
Since ${\rm Re}(T)$ is a linear combination of 4-cycle volumes which
corresponds to the invariant direction of non-perturbative superpotential,
while ${\rm Re}(\Phi^i)$ are other combinations corresponding to the
exponents of non-perturbative superpotential, such K\"ahler mixing between
$T$ and $\Phi^i$ is a generic feature of the moduli K\"ahler potential.

\begin{figure}[t]
\begin{center}
\begin{minipage}{15cm}
\centerline{
{\hspace*{0cm}\epsfig{figure=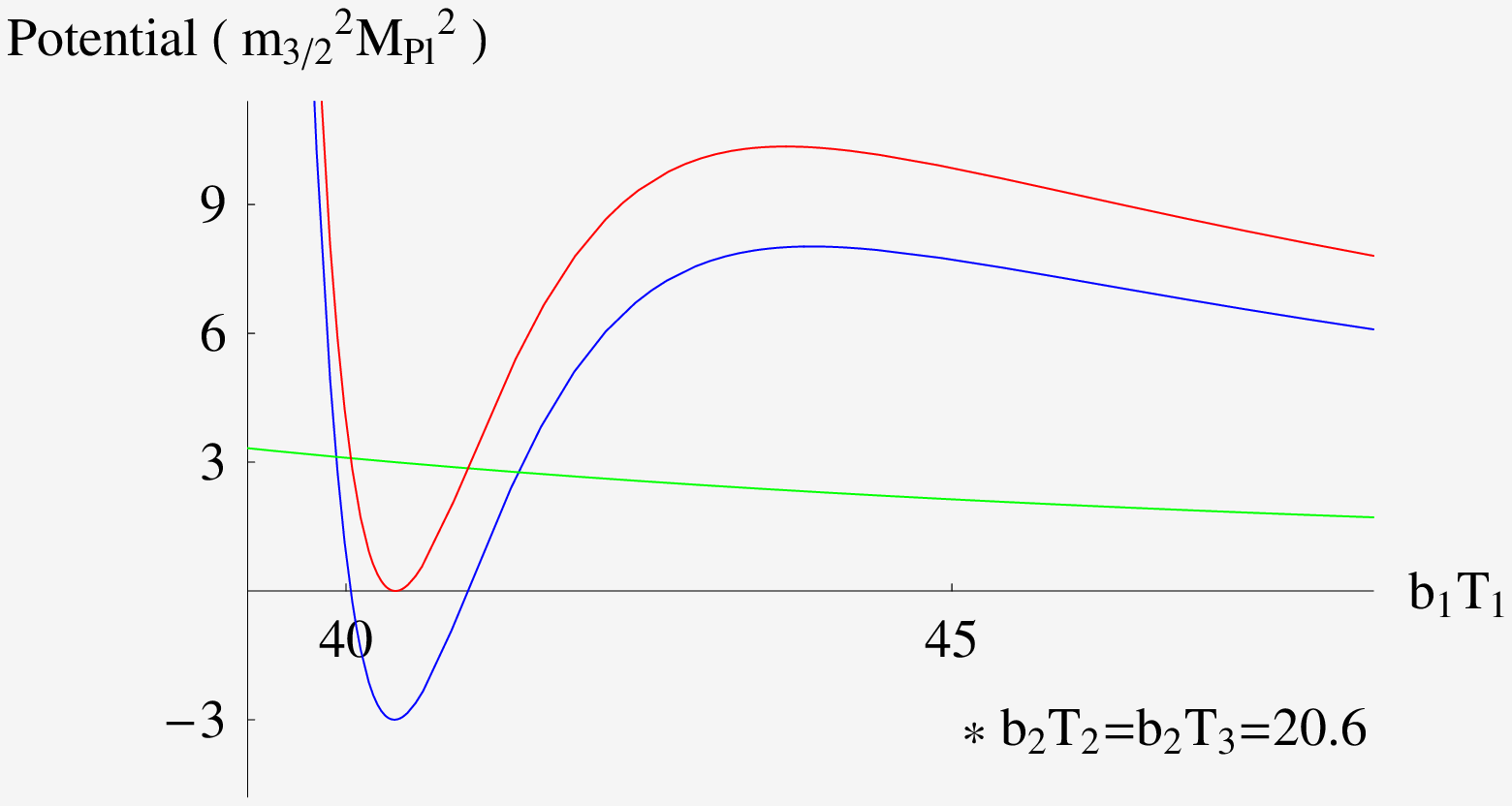,angle=0,width=7.3cm}}
{\hspace*{.2cm}\epsfig{figure=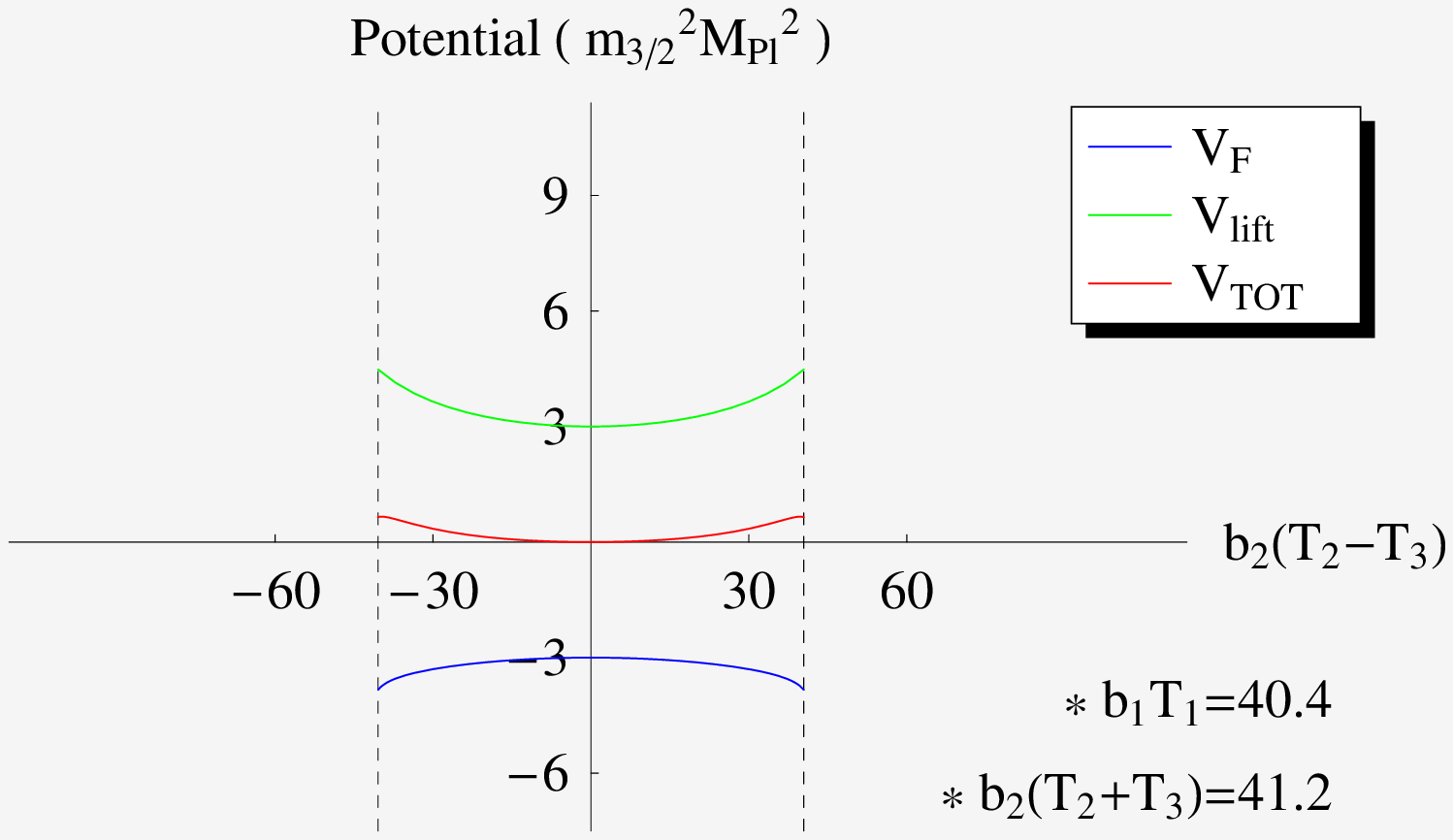,angle=0,width=7.3cm}}
}
\caption{
Moduli potential along $T_1$ and $T=T_2-T_3$ in unit of
$m_{3/2}^2M_{Pl}^2$ for the toy model (2.7).
$T_1$ is stabilized by
the KKLT superpotential giving $m_{T_1}\sim m_{3/2}\ln(M_{Pl}/m_{3/2})$,
while the saxion ${\rm Re}(T)$ is stabilized with
$m_s\simeq \sqrt{2}m_{3/2}$ by the sequestered uplifting potential.
\label{fig:potential}
}
\end{minipage}
\end{center}
\end{figure}

In Fig. 2, we show the behaviors of $V_F$, $V_{\rm lift}$ and
$V_{\rm TOT}$ along the KKLT K\"ahler modulus ${\rm Re}(T_1)$ and
the saxion ${\rm Re}(T)={\rm Re}(T_2-T_3)$ for the toy example
(\ref{toy}). Note that all 4-cycle volumes ${\rm Re}(T_I)$
($I=1,2,3$) have large positive vacuum values in the limit
$m_{3/2}\ll M_{Pl}$, justifying the leading order K\"ahler
potential in (\ref{toy}). The vanishing saxion vacuum value,
$\langle {\rm Re}(T)\rangle=0$, is a result of convention, and
does not cause any trouble. In the next section, we provide a more
detailed analysis of the moduli masses and the pattern of SUSY
breaking $F$-components in generic $U(1)_{PQ}$-invariant
generalization of KKLT set-up.

\section{Moduli masses, $F$-components and the axion scale}

To examine the moduli masses and the pattern of SUSY breaking $F$-components,
let us expand the total moduli potential $V_{\rm TOT}=V_F+V_{\rm lift}$ and
the moduli $F$-components $F^I$ around the supersymmetric moduli
configuration $\Phi^I_0=(T_0,\Phi^i_0)$ satisfying
\bea
D_iW(\vec{\Phi}_0)=0,\quad
D_TW(\vec{\Phi}_0)=W(\vec{\Phi}_0)\partial_TK(\vec{\Phi}_0)=0.
\eea
We then find
\bea
\label{expansion}
V_{\rm TOT} &=&
-3|m_{3/2}(\vec{\Phi}_0)|^2+V_{\rm
lift}(\vec{\Phi}_0)+\sqrt{2}\delta\phi^I\partial_IV_{\rm
lift}(\vec{\Phi}_0) \nonumber
\\
&+&
\frac{1}{4}\delta\phi^I\delta\phi^J(\partial_I\partial_{\bar{J}}
+\partial_{\bar{I}}\partial_J +\partial_I\partial_J
+\partial_{\bar{I}}\partial_{\bar{J}})V_{\rm TOT}
\nonumber \\
&+&
\frac{1}{4}\delta a^I\delta
a^J(\partial_I\partial_{\bar{J}}
+\partial_{\bar I}\partial_J-\partial_I\partial_J
-\partial_{\bar{I}}\partial_{\bar{J}})V_{\rm TOT}
+{\cal O}((\delta\Phi)^3),
\nonumber \\
&=&-3|m_{3/2}(\vec{\Phi}_0)|^2+V_{\rm
lift}(\vec{\Phi}_0)+\frac{2\sqrt{2}}{3}\partial_IK
(\vec{\Phi}_0)V_{\rm lift}(\vec{\Phi}_0)\delta\phi^I
\nonumber \\
&+&
\frac{1}{2}\left(
m^2_\phi\right)_{IJ}\delta\phi^I\delta\phi^J+\frac{1}{2}
\left(m_a^2\right)_{IJ}\delta
a^I\delta a^J+{\cal O}((\delta\Phi)^3),
\nonumber \\
F^I
&=&
-\frac{1}{\sqrt{2}}m^*_{3/2}K^{I\bar{J}}\Big[\,(G_{\bar{J}\bar{L}}
+K_{\bar{J}{L}})\delta
\phi^L-i(G_{\bar{J}\bar{L}}-K_{\bar{J}{L}})\delta a^L\,\Big],
\eea
where
\bea
\Phi^I-\Phi^I_0 \equiv
\frac{1}{\sqrt{2}}(\delta\phi^I + i\delta a^I)=
\frac{1}{\sqrt{2}}(\delta s+i\delta a, \delta\phi^i+i\delta a^i)
\eea
are the moduli and axion fluctuations from the supersymmetric configuration
$\vec{\Phi}_0$, and
\bea
&& K_{I\bar{J}}=\partial_I\partial_{\bar{J}}K, \quad
G_{IJ}=\partial_I\partial_J(K+\ln|W|^2),\nonumber \\
&&\Big(m_\phi^2\Big)_{IJ}=|m_{3/2}|^2 \left(
G_{IL}G_{\bar{J}\bar{M}}K^{L\bar{M}}-
G_{IJ}-2K_{I\bar{J}}\right)\nonumber \\
&&\qquad \qquad +\,\frac{4}{3}\Big(
K_{I\bar{J}}+\frac{2}{3}\partial_IK\partial_{\bar{J}}K\Big)\,V_{\rm
lift},
\nonumber\\
&&\Big(m_a^2\Big)_{IJ}=|m_{3/2}|^2(G_{IL}G_{\bar{J}\bar{M}}K^{L\bar{M}}+
G_{IJ}-2K_{I\bar{J}})
\eea for
$\partial_I=\partial/\partial\Phi^I$ and
$\partial_{\bar{I}}=\partial/\partial\Phi^{I*}$.
Here we are using that
$K=K(\Phi^I+\Phi^{I*})$, $V_{\rm lift}(\Phi^I+\Phi^{I*})$,
$\vec{\Phi}_0$ is a stationary point of $V_F$, and the effective SUGRA
(\ref{Action}) preserves CP as a consequence of $U(1)_i$ and $U(1)_{PQ}$
which assure that $w_0$ and $A_i$ in $W$ can be chosen to be real and all
derivatives of $K$ and $V_{\rm lift}$ are automatically real \cite{choi2}.
The kinetic terms of moduli and axion fluctuations are given by
\bea
\label{kinetic}
{\cal L}_{\rm kin}= \frac{1}{2}K_{I\bar{J}}(\vec\Phi_0)
\Big[\,
\partial_\mu\delta\phi^I\partial^\mu\delta\phi^J
+ \partial_\mu\delta a^I\partial^\mu\delta
a^J\Big].
\eea

The SUSY condition $D_iW=0$ gives
\bea
\label{phi}
\Phi^i_0 =
\frac{1}{b_i}\ln\left(\frac{A_i(b_i-\partial_iK)}{w_0\partial_iK}\right)
=\frac{\ln({M_{Pl}}/{m_{3/2}})}{b_i}
\left[1+{\cal O}\left(
\frac{1}{\ln(M_{Pl}/m_{3/2})}\right)\right],
\eea
where we assumed that $A_i$ are of the order of unity, while
$w_0\sim m_{3/2}$ is hierarchically small to get the low energy SUSY at
final stage\footnote{
Such a small $w_0$ might be achieved by tuning the 3-form fluxes, or by
non-perturbative effects if 3-form fluxes preserve a discrete $R$-symmetry,
e.g. $D3$ brane gaugino condensation which would give $w_0\sim e^{-bS_0}$,
where $S_0$ is the vacuum value of the massive Type IIB dilaton.
}.
In fact, the little hierarchy factor $\ln(M_{Pl}/m_{3/2})$ allows a
perturbative expansion in powers of
\bea
\epsilon \equiv
\frac{1}{\ln(M_{Pl}/m_{3/2})}\sim \frac{1}{4\pi^2}.
\eea
Note that $b_i\Phi^i_0$ (no summation over $i$) have {\it universal} values
at leading order in $\epsilon$.
In the normalization convention of K\"ahler moduli for which
${\rm Re}(\Phi^i_0)\sim 1$, we have $b_i\sim \ln(M_{Pl}/m_{3/2})$,
while $K$ and their derivatives are generically of order unity.

The uplifting potential shifts the moduli vacuum values as
\bea
\delta\phi^I=
-\frac{2\sqrt{2}}{3}\Big(m_\phi^{2}\Big)^{-1}_{IJ}(\partial_JK)
V_{\rm lift},
\qquad \delta a^I=0.
\eea
As we will see, the moduli shifts $\delta\phi^I$ are all of
${\cal O}(\epsilon^2)$.
Although tiny, this vacuum shift is the origin of nonzero vacuum values of
$F^I$ which were vanishing before the shift.
On the other hand, the moduli masses at true vacuum
$\langle \vec{\Phi}\rangle=\vec{\Phi}_0+\delta\vec{\Phi}$ can be approximated
well by the values at $\vec{\Phi}_0$ since the vacuum shift
$\delta\vec\Phi={\cal O}(\epsilon^2)$ gives a small correction to moduli
masses.

For $\delta\vec{\Phi}={\cal O}(\epsilon^2)$, the condition of vanishing
cosmological constant requires
\bea
V_{\rm lift}(\vec{\Phi}_0)\simeq 3|m_{3/2}|^2.
\eea
It is also straightforward to see that the moduli and axion masses are given
by
\bea
\label{modulimass}
&& \Big(m_\phi^2\Big)_{ij}={\cal O}
\left( (m_{3/2}\ln(M_{Pl}/m_{3/2}))^2\right) ,\quad
\Big(m_\phi^2\Big)_{TI}=2m_{3/2}^2\partial_T\partial_IK,
\nonumber \\
&& \Big(m_a^2\Big)_{ij}= {\cal O}
\left((m_{3/2}\ln(M_{Pl}/m_{3/2}))^2\right),
\quad\Big(m_a^2\Big)_{TI}=0.
\eea
From this, one can easily finds
$\delta\phi^I=(\delta s, \delta\phi^i)$ are all of the order of
$\epsilon^2$.
The reason for $\delta\phi^i={\cal O}(\epsilon^2)$ is that $\phi^i$ are heavy
as $m^2_\phi\sim m_{3/2}^2/\epsilon^2$.
On the other hand, $m_s^2\sim m_{3/2}^2$, thus the reason for
$\delta s={\cal O}(\epsilon^2)$ is different.
Since both $\partial_TV_F$ and $\partial_TV_{\rm lift}$ are vanishing at
$\vec{\Phi}_0$, the uplifting potential does not directly induce a saxion
tadpole. Rather, the saxion tadpole is induced by $\delta\phi^i$ through the
K\"ahler mixing with $\phi^i$.
Explicitly, we find
\bea
\delta\phi^i &=&
-\frac{2\sqrt{2}}{b_i\partial_iK}\sum_j
\frac{1}{b_j}\left( \partial_i\partial_{\bar j}K
- \frac{(\partial_i\partial_{\bar T}K)\partial_T\partial_{\bar j}K}
{\partial_T\partial_{\bar T}K} \right),
\nonumber \\
\delta s &=&
-\frac{1}{\partial_T\partial_{\bar{T}}K}\sum_i\delta\phi^i
\partial_T\partial_{\bar{i}}K,
\eea
for which
\bea
\label{fcomponent}
F^i &=&
\frac{2m_{3/2}}{b_i}\,=\,(\Phi^i+\Phi^{i*})
\frac{m_{3/2}}{\ln(M_{Pl}/m_{3/2})}, \nonumber \\
F^T &=& -\sum_i
\frac{\partial_i\partial_{\bar{T}}K}{\partial_T\partial_{\bar{T}}K}F^i
\,=\,
-\frac{m_{3/2}}{\ln(M_{Pl}/m_{3/2})}\sum_i
\frac{(\Phi^i+\Phi^{i*})\partial_i\partial_{\bar{T}}K}
{\partial_T\partial_{\bar{T}}K},
\eea
at leading order in $\epsilon$.
Note that ${\rm Re}(\Phi^i)$ are defined as the exponents of
non-perturbative terms in KKLT superpotential
$W=w_0+\sum_i A_ie^{-b_i\Phi^i}$, thus correspond to the linear combinations
of the 4-cycle volume moduli which obtain large positive vacuum values
$\langle b_i\Phi^i\rangle\simeq \ln(M_{Pl}/m_{3/2})\gg 1$ by
non-perturbative superpotential.
On the other hand, ${\rm Re}(T)$ is a linear combination which can have any
sign of vacuum value, even a vanishing vacuum value in some case such as the
model (\ref{toy}).
All of the above results are valid independently of the sign of the vacuum
value of $T$.
The $F$-components $F^i/(\Phi^i+\Phi^{i*})$ are universal at leading order
in $\epsilon$ independently of the detailed form of the moduli K\"ahler
potential, which is a characteristic  feature of the KKLT-type moduli
stabilization \cite{choi3}.
On the other hand, $F^T$ appears to depend on the detailed form of $K$,
particularly on $\partial_i\partial_{\bar{T}}K$.
As we will see, if $K$ takes a no-scale form, $F^T/(T+T^*)$ also becomes
same as the universal $F^i/(\Phi^i+\Phi^{i*})$, thereby all moduli
$F$-components (divided by moduli vacuum value) have universal values.

Combined with the moduli-axion kinetic term (\ref{kinetic}), the mass
matrices (\ref{modulimass}) give the following pattern of moduli and axion
mass eigenvalues:
\bea
\label{moduliaxionmass}
&&
m_{\phi_i}\simeq m_{a_i} \sim
m_{3/2}\ln(M_{Pl}/m_{3/2}),
\nonumber \\
&&
m_s = \sqrt{2}m_{3/2},\quad m_a=0,
\eea
where the mass eigenstate saxion and axion are mostly ${\rm Re}(T)$ and
${\rm Im}(T)$, respectively.
The SUSY-breaking $F$-components of all moduli are of the order of
$m_{3/2}/4\pi^2$:
\bea
\label{susybreaking}
\frac{F^I}{\Phi^I+\Phi^{I*}} \sim
\frac{m_{3/2}}{\ln(M_{Pl}/m_{3/2})} \sim
\frac{m_{3/2}}{4\pi^2}.
\eea
As a result, the soft SUSY breaking terms of visible fields receive
comparable contributions from moduli mediation and anomaly mediation
\cite{choi1}, leading to the mirage mediation pattern of superparticle
masses discussed in \cite{mirage,mirage1,mirage2,mirage3}.
We also find the modulino and axino masses are given by
\bea
m_{\tilde{\phi}_i}\sim m_{3/2}\ln(M_{Pl}/m_{3/2}), \quad
m_{\tilde{a}}=m_{3/2}.
\eea
Note that the helicity 1/2 component of gravitino mostly comes from the
Goldstino localized on the SUSY breaking brane, not from the K\"ahler
modulino/axino.

We stress that the above results of moduli/modulino masses and
$F$-components are obtained for generic effective SUGRA under the
assumptions of
(i) the axionic shift symmetries
\bea
\label{axionicshift}
U_{PQ}: T\rightarrow T+i\beta,\quad U(1)_i:\Phi^i \rightarrow
\Phi^i+i\beta_i,
\eea
broken dominantly by non-perturbative superpotential except for $U(1)_{PQ}$,
and
(ii) a sequestered uplifting potential.
In compactified string theory, the pseudoscalar components of K\"ahler
moduli correspond to the zero modes of antisymmetric tensor gauge field,
and thus the axinonic shift symmetries broken only by non-perturbative
effects are generic feature of 4D effective theory.
Also, as we have noticed, a sequestered uplifting potential is the most
plausible form of uplifting mechanism in string compactification with warped
throat.
It arises from generic SUSY-breaking brane stabilized at the IR end of
warped throat.
Thus the moduli mass pattern (\ref{moduliaxionmass}) and the SUSY-breaking
$F$-components (\ref{susybreaking}) can be considered as a quite robust
prediction of $U(1)_{PQ}$-invariant generalization of KKLT moduli
stabilization  giving the QCD axion.

Now, let us note that the K\"ahler moduli $F$-components become
{\it universal} for the no-scale K\"ahler potential satisfying
\bea
\label{noscale}
\sum_JK^{I\bar{J}}\partial_{\bar{J}}K=-(\Phi^I+\Phi^{I*}).
\eea
Indeed, at leading order in $\alpha^\prime$ and string loop expansion,
the K\"ahler potential of K\"ahler moduli in Type IIB string compactification
takes the no-scale form.
Then for SUSY configuration satisfying $\partial_TK=0$, the above relation
gives
\bea
-\partial_{\bar T}K = \sum_I(\Phi^I+\Phi^{I*})K_{I\bar{T}}
= (T+T^*)\partial_T\partial_{\bar{T}}K
+ \sum_i(\Phi^i+\Phi^{i*})\partial_i\partial_{\bar{T}}K=0
\eea
Applying this to (\ref{fcomponent}), we find
\bea
\label{universality}
\frac{F^T}{T+T^*}=
\frac{F^i}{\Phi^i+\Phi^{i*}}=\frac{m_{3/2}}{\ln(M_{Pl}/m_{3/2})},
\eea
i.e. all K\"ahler moduli including  $T$ have universal values of
$F^I/\Phi^I$ although $T$ and $\Phi^i$ are stabilized by different
mechanisms.
As we will see, this universality of $F^I/\Phi^I$ has an interesting
implication for the flavor conservation of soft terms.

Much of the low energy properties of the QCD axion is determined by the
axion decay constant $v_{\rm PQ}$ which is defined through the effective
coupling between the axion and the gluon anomaly:
\bea
{\cal L}_{\rm eff} = \frac{1}{2}\partial_\mu a\partial^\mu
a-\frac{1}{4g_{QCD}^2}F^{a\mu\nu}F^a_{\mu\nu}+
\frac{1}{32\pi^2}\frac{a}{v_{\rm PQ}}F^{a\mu\nu}\tilde{F}^a_{\mu\nu},
\eea
where
$\tilde{F}^{a\mu\nu}=\frac{1}{2}\epsilon^{\mu\nu\rho\sigma}F^a_{\rho\sigma}$.
Using that $a$ is contained mostly in ${\rm Im}(T)$, for the QCD gauge
kinetic function $f_{\rm QCD}$ which takes the form of (\ref{gaugekinetic}),
we find
\bea
v_{\rm PQ} &=&
\frac{M_{Pl}}{8\pi^2}
\,\frac{(\partial_T\partial_{\bar{T}}K)^{1/2}}
{\partial_T [\ln{\rm Re}(f_{\rm QCD})]},
\eea
where $M_{Pl}\simeq 2\times 10^{18}$ GeV.
As the above result shows, the precise value of $v_{\rm PQ}$ is somewhat
model-dependent, but generically around $10^{16}$ GeV.

Such a large value of $v_{\rm PQ}$ might cause the cosmological problem that
the cosmological axion density produced by initial misalignment overcloses
the Universe \cite{axioncos}.
Interestingly, for the moduli stabilization scenario under consideration,
this cosmological axion problem can be significantly ameliorated by the late
decay of saxion \cite{saxiondecay} which has a right mass to decay right
before the big-bang nucleosynthesis (BBN) for the most interesting case that
the visible sector superparticle masses are of the order of the weak scale.
As will be discussed in the next section, the soft SUSY breaking masses of
visible fields are given by
$m_{\rm soft}\sim \frac{m_{3/2}}{\ln(M_{Pl}/m_{3/2})}$.
As a result, the saxion mass $m_s\simeq \sqrt{2}m_{3/2}\sim 50$ TeV and
$m_{\phi_i}\simeq m_{a_i}\sim m_{3/2}\ln(M_{Pl}/m_{3/2})\sim 10^3$
TeV for $m_{\rm soft}\sim 1$ TeV.
On the other hand, moduli (including the saxion) couplings to visible fields
are suppressed by $1/M_{Pl}\sim 1/8\pi^2 v_{\rm PQ}$, giving the reheat
temperature after saxion decay $T_{RH} \sim 6$ MeV for $m_s\sim 50$ TeV.
If the early Universe before the saxion decay were dominated by the coherent
oscillation of saxion field, such late decay of saxions dilutes the axion
density as well as the potentially dangerous primordial gravitinos, therefore
might allow the model to avoid the cosmological gravitino, moduli and axion
problems \cite{saxiondecay}.
In particular, in this scenario, the QCD axion can be a good dark matter
candidate under a mild assumption on the initial axion misalignment although
the axion decay constant is around the GUT scale.
A detailed analysis of the moduli and axion cosmology in the generalized
KKLT set up with QCD axion will be presented elsewhere \cite{choi4}.

\section{Flavor and CP conserving soft terms}

In this section, we discuss the soft terms in more detail,
focusing on the CP and flavor issues.
To this end, let us include the visible gauge and matter superfields in the
effective SUGRA action:
\bea
\label{superspace}
&&\int d^4\theta CC^*\left[
-3\exp\left\{-\frac{1}{3}
\Big(K_0+Z_pQ^pQ^{p*}\Big)\right\}
-CC^*\Lambda^2\bar{\Lambda}^2e^{4A}{\cal P}_0\right]
\nonumber \\
&&+\, \left[\,\int d^2\theta \left\{\frac{1}{4}f_a
W^{a\alpha}W^a_\alpha+ C^3\Big(
w_0+\sum_iA_ie^{-b_i\Phi^i}+\frac{1}{6}\lambda_{pqr}Q^pQ^qQ^r\Big)\right\}
+{\rm h.c.}\,\right],
\eea
where the axionic shift symmetries (\ref{axionicshift}) require that moduli
K\"ahler potential $K_0$, matter K\"ahler metric $Z_p$, and the holomorphic
gauge kinetic functions are given by
\bea
\label{invariantfunction}
K_0 &=& K_0(\Phi^I+\Phi^{I*}),\quad
Z_p\,=\,Z_p(\Phi^I+\Phi^{I*}),
\nonumber \\
f_a &=& \sum_I c_I\Phi^I = c_TT+\sum_ic_{i}\Phi^i, \eea where
$c_I$ are real constants with nonzero $c_T$. The soft SUSY
breaking terms of canonically normalized visible fields can be
written as \bea {\cal L}_{\rm
soft}&=&-\frac{1}{2}M_a\lambda^a\lambda^a-\frac{1}{2}m_r^2|\tilde{Q}^r|^2
-\frac{1}{6}A_{pqr}y_{pqr}\tilde{Q}^p\tilde{Q}^q\tilde{Q}^r +{\rm
h.c.}, \eea where $\lambda^a$ are gauginos, $\tilde{Q}^r$ are the
scalar component of $Q^r$ and $y_{pqr}$ are the canonically
normalized Yukawa couplings: \bea y_{pqr} =
\frac{\lambda_{pqr}}{\sqrt{e^{-K_0}Z_pZ_qZ_r}}. \eea For
$F^I/\Phi^I\sim m_{3/2}/4\pi^2$, the soft parameters at energy
scale just below $M_{GUT}\sim 2\times 10^{16}$ GeV are determined
by the modulus-mediated and anomaly-mediated contributions which
are comparable to each other. One then finds \cite{choi1}
\begin{eqnarray}
\label{soft1}
M_a &=& \tilde{M}_a
+\frac{m_{3/2}}{16\pi^2}\,b_ag_a^2,
\nonumber \\
A_{pqr} &=&\tilde{A}_{pqr}-
\frac{m_{3/2}}{16\pi^2}\,(\gamma_p+\gamma_q+\gamma_r),
\nonumber \\
m_r^2 &=& \tilde{m}_r^2-\frac{1}{16\pi^2}(m_{3/2}^*\Theta_r
+ m_{3/2}\Theta^*_r)
-\left|\frac{m_{3/2}}{16\pi^2}\right|^2\dot{\gamma}_i,
\label{eq:bc}
\end{eqnarray}
where the moduli-mediated contributions are given by
\bea
\label{tmediation}
\tilde{M}_a &=&\sum_IF^I\partial_I\ln{\rm Re}(f_a),
\nonumber \\
\tilde{m}_r^2 &=&
-\sum_{IJ}F^IF^{\bar{J}}\partial_I\partial_{\bar{J}}
\ln(e^{-K_0/3}Z_r),
\nonumber \\
\tilde{A}_{pqr}&=& -\sum_IF^I\partial_I\ln
\left(\frac{\lambda_{pqr}}{e^{-K_0}Z_pZ_qZ_r}\right)
\,=\, \sum_IF^I\partial_I\ln(e^{-K_0}Z_pZ_qZ_r).
\eea
Here we have used that the holomorphic Yukawa couplings $\lambda_{pqr}$ are
independent of $\Phi^I$ as a consequence of the axionic shift symmetries.
The one-loop beta function coefficient $b_a$, the anomalous dimension
$\gamma_p$ and its derivative $\dot{\gamma}_p$, and $\Theta_p$ are defined
as
\bea
b_a &=& -3{\rm tr}\left(T_a^2({\rm Adj})\right)
        +\sum_p {\rm tr}\left(T^2_a(Q^p)\right),
\nonumber \\
\gamma_p &=& 2\sum_aC^a_2(Q^p)g_a^2-\frac{1}{2}\sum_{qr}|y_{pqr}|^2,
\nonumber \\
\dot{\gamma}_p&=&8\pi^2\frac{d\gamma_p}{d\ln\mu},
\nonumber \\
\Theta_p &=&
2\sum_aC^a_2(Q^p)g_a^2\tilde{M}_a-\frac{1}{2}\sum_{qr}|y_{pqr}|^2
\tilde{A}_{pqr},
\eea
where the quadratic Casimir $C^a_2(Q^p)=(N^2-1)/2N$ for a fundamental
representation $Q^p$ of the gauge group $SU(N)$,
$C_2^a(Q^p)=q_p^2$ for the $U(1)$ charge $q_p$ of $Q^p$, and
$\omega_{pq}=\sum_{rs}y_{prs}y^*_{qrs}$ is assumed to be diagonal.

As was noticed before \cite{choi2}, the axionic shift symmetries
(\ref{axionicshift}) assure that the above soft terms preserve CP.
To see this, let us first note that $\partial_IK_0$,
$\partial_IZ_p$, $\partial_I V_{\rm lift}$,
$\partial_I\lambda_{pqr}$, and  $\partial_If_a$ are all real as a consequence
of the axionic shift symmetries. Combined with $U(1)_R$ transformation,
the axionic shift symmetries also allow that $w_0$ and $A_i$ in the moduli
superpotential are chosen to be real without loss of generality, leading to
real $m_{3/2}$ and $F^I$.
Obviously then all soft parameters are real, thus preserve CP
{\it independently of} the detailed forms of the moduli K\"ahler potential,
matter K\"ahler metric and the gauge kinetic functions as long as they
respect the axionic shift symmetries.

Recently, it has been noticed that $e^{-K_0/3}Z_p$ have a definite
scaling property under the overall rescaling of the CY metric,
$g_{mn}\rightarrow \lambda g_{mn}$, at leading order in
$\alpha^\prime$ and string loop expansion \cite{quevedo1}. Under
this rescaling of metric, the CY volume and K\"ahler moduli
transform as \bea V_{CY}\rightarrow \lambda^3 V_{CY}, \qquad
\Phi^I\rightarrow \lambda^2\Phi^I, \eea  while the IIB dilaton and
complex structure moduli do not transform. The normalized
wavefunction of matter zero mode $Q_p$ transforms as \bea \psi_p
\rightarrow \lambda^{-d_p/4}\psi_p \eea under the metric
rescaling, where $d_p$ is the internal dimension of the subspace
$\sigma_p$ over which $Q_p$ can propagate. The physical Yukawa
couplings are then given by the integral of matter wavefunctions
over a subspace $\sigma_{pqr}$ of the intersection of $\sigma_p$,
$\sigma_q$ and $\sigma_r$: \bea y_{pqr}=\int_{\sigma_{pqr}}
dx^{d_{pqr}}\sqrt{g}\psi_p\psi_q\psi_r \eea which transforms as
\bea y_{pqr} \rightarrow
\lambda^{(2d_{pqr}-d_q-d_q-d_r)/4}y_{pqr}, \eea where $d_{pqr}$ is
the dimension of $\sigma_{pqr}$. On the other hand, the
holomorphic Yukawa couplings $\lambda_{pqr}$ in 4D effective SUGRA
do not transform under the metric rescaling since they are
independent of K\"ahler moduli due to the axionic shift symmetries
(\ref{axionicshift}). Then, to match with the scaling property of
$y_{pqr}$ which is given by
$y_{pqr}=\lambda_{pqr}/\sqrt{e^{-K_0}Z_pZ_qZ_r}$ in 4D effective
SUGRA, the matter K\"ahler metric $Z_p$ should transform as \bea
Z_p(\lambda^2(\Phi^I+\Phi^{I*}))=
\lambda^{2(n_p-1)}Z_p(\Phi^I+\Phi^{I*}), \eea where the scaling
weights $n_p$ satisfy \bea 4(n_p+n_q+n_r)=d_p+d_q+d_r-2d_{pqr}
\eea for the combinations of $Q_p$ with nonzero Yukawa coupling.
Here we have used that the leading order K\'ahler potential of
K\"ahler moduli is given by $K_0=-2\ln(V_{CY})$.

Combined with the universality of $F^I/\Phi^I$  obtained for a
no-scale form of moduli K\"ahler potential, the above scaling
property of the matter K\"ahler metric assures that soft terms
derived from our moduli stabilization set-up preserve flavor. For
the universal $F$-components: \bea
\frac{F^I}{\Phi^I+\Phi^{I*}}\,=\,\frac{m_{3/2}}{\ln(M_{Pl}/m_{3/2})}
\,\equiv\, M_0, \eea the K\"ahler potential and gauge kinetic
functions of (\ref{invariantfunction}) give \bea
\label{modulimediation1} \tilde{M}_a &=&
M_0\sum_I(\Phi^I+\Phi^{I*})\partial_I\ln{\rm Re}(f_a)\,=\, M_0,
\nonumber \\
\tilde{m}^2_r &=&
-M_0^2\sum_{IJ}(\Phi^I+\Phi^{I*})(\Phi^J+\Phi^{J*})
\partial_I\partial_{\bar{J}}\ln(e^{-K_0/3}Z_r),
\nonumber \\
\tilde{A}_{pqr} &=& M_0
\sum_I(\Phi^I+\Phi^{I*})\partial_I\ln(e^{-K_0}Z_pZ_qZ_r). \eea
Then the scaling property of $e^{-K_0/3}Z_p$ leads to \bea
\label{modulimediation2} \tilde{m}^2_r\,=\, n_rM_0^2, \quad
\tilde{A}_{pqr}\,=\, (n_p+n_q+n_r)M_0, \eea i.e. the
moduli-mediated soft scalar masses and $A$-parameters are
determined simply by the matter scaling weights. It is highly
plausible that matter fields with the same gauge quantum numbers
have a common geometric origin, therefore have the same scaling
weights. Then the soft masses take a phenomenologically desirable
{\it flavor-blind} form independently of the detailed form of the
matter and moduli K\"ahler potential at leading order in
$\alpha^\prime$ and string loop expansion.

Taking into account the 1-loop RG evolution, the soft masses of
(\ref{soft1}) at $M_{GUT}\sim 2\times 10^{16}$ GeV lead to low
energy soft masses described by the mirage messenger scale
\cite{mirage}:
\bea
M_{\rm mir} \,\sim\,
\frac{M_{GUT}}{(M_{Pl}/m_{3/2})^{1/2}}\,\sim\, 3\times 10^9\,\,
{\rm GeV}.
\eea
The low energy gaugino masses are given by
\bea
\label{lowgaugino}
M_a(\mu) = M_0\left[\,
1-\frac{1}{8\pi^2}\,b_ag_a^2(\mu)
\ln\left(\frac{M_{\rm mir}}{\mu}\right)\,\right] =
\frac{g_a^2(\mu)}{g_a^2(M_{\rm mir})}M_0,
\eea
showing that the gaugino masses are unified at $M_{\rm mir}$,
while the gauge couplings are unified at $M_{GUT}$.
The low energy values of $A_{pqr}$ and $m_r^2$ generically depend
on the associated Yukawa couplings $y_{pqr}$. However if $y_{pqr}$
are small enough, e.g. the case of the first and second
generations of quarks and leptons, or
\bea
\label{condition1}
n_p+n_q+n_r=1 \quad \mbox{for} \quad y_{pqr}\sim 1,
\eea
their low energy values are given by \cite{mirage}
\bea
\label{lowsfermion}
A_{pqr}(\mu)&=& M_0\left[\,n_p+n_q+n_r+
\frac{1}{8\pi^2}(\gamma_p(\mu)
+\gamma_q(\mu)+\gamma_r(\mu))
\ln\left(\frac{M_{\rm mir}}{\mu}\right)\,\right], \nonumber \\
m_r^2(\mu)&=&M_0^2\left[\,n_r-\frac{1}{8\pi^2}Y_r\left(
\sum_pn_pY_p\right)g^2_Y(\mu)\ln\left(\frac{M_{GUT}}{\mu}\right)\right.
\nonumber \\
&&
\hspace{1cm} + \left.\frac{1}{4\pi^2}\left\{
\gamma_r(\mu)-\frac{1}{2}\frac{d\gamma_r(\mu)}{d\ln\mu}\ln\left(
\frac{M_{\rm mir}}{\mu}\right)\right\}\ln\left( \frac{M_{\rm
mir}}{\mu}\right)\,\right],
\eea
where $Y_p$ is the $U(1)_Y$ charge of $Q^p$.
In this case, the $A$-parameters and sfermion masses at $M_{\rm mir}$
are (approximately) same\footnote{ Note
that $\sum_p n_p Y_p=0$ if the scaling weights $n_p$ obey the
$SU(5)$ relation. Even when $\sum_pn_pY_p\neq 0$, the part of
$m_r^2(M_{\rm mir})$ due to nonzero $\sum_pn_pY_p$ is numerically
negligible.} as the moduli-mediated contributions at $M_{GUT}$. If
the squarks and sleptons have common scaling weight, which is a
rather plausible possibility, the squark and slepton masses appear
to be unified at $M_{\rm mir}$ for the case that either the
associated Yukawa couplings are small or  the scaling weights obey
the condition (\ref{condition1}). We note that (\ref{condition1})
is obtained when  $d_p+d_q+d_r-2d_{pqr}=4$,  for instance when
$y_{pqr}$ is given by an integral of the quark/lepton and Higgs
wavefunctions over 4-cycle, for which $d_p=d_q=d_r=d_{pqr}=4$, or
when $y_{pqr}$ is given by an integral over 2-cycle of the
quarks/leptons wavefunctions defined on 2-cycle  and the Higgs
wavefunction defined on 4-cycle, for which $d_{pqr}=2$, $d_{\rm
quark}=d_{\rm lepton}=2$, and $d_{\rm Higgs}=4$.

So far, we have discussed the soft terms in generic
$U(1)_{PQ}$-invariant generalization of KKLT moduli stabilization
which can give the QCD axion solving the strong CP problem. The
soft terms preserve CP due to the axionic shift symmetries broken
by non-perturbative superpotential \cite{choi2}. At leading order
in $\alpha^\prime$ and string loop expansion, the moduli
$F$-components $F^I/\Phi^I$ have universal vacuum values and also
the matter K\"ahler metrics have a definite scaling property under
the overall rescaling of CY metric. As a consequence, the
moduli-mediated contributions to soft terms take a highly
predictive form, i.e. the moduli-mediated gaugino masses at
$M_{GUT}$ are given by $\tilde{M}_a=M_0\equiv
m_{3/2}/\ln(M_{Pl}/m_{3/2})$ and the moduli-mediated
$A$-parameters and sfermion masss at $M_{GUT}$ are simply
determined by the scaling weights as (\ref{modulimediation2}),
independently of the detailed forms of the  moduli K\"ahler
potential and matter K\"ahler metric. Since the matter fields with
the same gauge quantum numbers are expected to have common scaling
weight, the soft terms in our moduli stabilization set-up
naturally preserve flavor  at leading order in $\alpha^\prime$ and
string loop expansion . If the higher order
corrections\footnote{An example of potentially important higher
order correction would be the effect of magnetic flux ${\cal F}$
on $D7$ brane wrapping a 4-cycle $\sigma$, whose strength is
controlled by $\frac{\alpha^{\prime 2}}{8\pi^2}\int_\sigma {\cal
F}\wedge{\cal F}$.}
 in underlying string compactification can be made to be small enough,
this flavor-universality at leading order approximation would
ensure that the model naturally passes the constraints from low
energy flavor violation.

\section{Conclusion}

The QCD axion provides an attractive solution to the strong CP problem.
As it contains numerous axions, string theory is perhaps the most plausible
framework to give the QCD axion.
In supersymmetric compactification, the QCD axion accompanies its scalar
partner, the saxion, which should be stabilized while keeping the QCD axion
as a flat direction until the low energy QCD instanton effects are taken
into account.
In this paper, we show that a simple generalization of KKLT moduli
stabilization  with the number of K\"ahler moduli $h_{1,1}>1$ provides such
saxion stabilization set-up.

Quite interestingly, although some details of  moduli stabilization are
different from the original KKLT scenario, the resulting soft SUSY breaking
terms still receive comparable contributions from moduli mediation
(including the saxion mediation) and anomaly mediation,
therefore take the mirage mediation pattern with
$m_{\rm soft}\sim \frac{m_{3/2}}{\ln(M_{Pl}/m_{3/2})}$ as in the KKLT case.
Furthermore, the soft terms naturally preserve flavor and CP as a
consequence of approximate scaling and axionic-shift symmetries of the
underlying string compactification, independently of the detailed forms of
moduli K\"ahler potential and matter K\"ahler metric.
As for the moduli spectrum, saxion has a mass $m_s\simeq \sqrt{2}m_{3/2}$,
while other moduli (except for the QCD axion) have a mass of the order of
$m_{3/2}\ln(M_{Pl}/m_{3/2})$ or heavier,
independently of the detailed form of the moduli K\"ahler potential.
This pattern of moduli masses might avoid the cosmological gravitino,
moduli and axion problems for the most interesting case that
$m_{\rm soft}\sim 1$ TeV since the saxion has a right mass to decay right
before the big-bang nucleosynthesis.
In particular, it might allow the QCD axion to be a good dark matter
candidate under a mild assumption on the initial axion misalignment although
the axion decay constant is near the GUT scale.

\vspace{1cm}
\vspace{5mm} \noindent{\large\bf Acknowledgments}
\vspace{5mm}

This work is supported by the KRF Grant funded by the Korean
Government (KRF-2005-201-C00006), the KOSEF Grant (KOSEF
R01-2005-000-10404-0), and the Center for High Energy Physics of
Kyungpook National University.

\end{document}